\documentclass{article}

\usepackage{graphicx}
\usepackage{amssymb}

\usepackage{titlesec}
\titleformat{\section}{\large\bfseries}{\thesection}{1em}{}
\titleformat{\subsection}{\normalsize\bfseries}{\thesubsection}{1em}{}

\usepackage[font=small,labelfont=bf, labelsep=period]{caption}
\begin{document}

\thispagestyle{empty}

\begin{center}
  {\Large \textbf{Hidden Breakpoints in Genome Alignments}}

  \vspace{2em}
  Birte Kehr$^{1,2}$, Knut Reinert$^{1}$, Aaron E. Darling$^{3}$

  \vspace{1em}
  \small
  $^1$ Department of Computer Science, Freie Universit\"at Berlin, Takustr.\,9,~14195~Berlin,~Germany\\
  $^2$ Max Planck Institute for Molecular Genetics, Ihnestr.\,63-73, 14195 Berlin, Germany\\
  $^3$ Genome Center, University of California-Davis, Davis, CA 95616

  \vspace{3em}
  \begin{minipage}{0.85\textwidth}
  \small
  \paragraph{Abstract.}
During the course of evolution, an organism's genome can undergo changes that affect the large-scale structure of the genome.
These changes include gene gain, loss, duplication, chromosome fusion, fission, and rearrangement.
When gene gain and loss occurs in addition to other types of rearrangement, breakpoints of rearrangement can exist that are only detectable by comparison of three or more genomes.
An arbitrarily large number of these ``hidden'' breakpoints can exist among genomes that exhibit no rearrangements in pairwise comparisons.

We present an extension of the multichromosomal breakpoint median problem to genomes that have undergone gene gain and loss.
We then demonstrate that the median distance among three genomes can be used to calculate a lower bound on the number of hidden breakpoints present.
We provide an implementation of this calculation including the median distance, along with some practical improvements on the time complexity of the underlying algorithm.

We apply our approach to measure the abundance of hidden breakpoints in simulated data sets under a wide range of evolutionary scenarios.
We demonstrate that in simulations the hidden breakpoint counts depend strongly on relative rates of inversion and gene gain/loss.
Finally we apply current multiple genome aligners to the simulated genomes, and show that all aligners introduce a high degree of error in hidden breakpoint counts, and that this error grows with evolutionary distance in the simulation.
Our results suggest that hidden breakpoint error may be pervasive in genome alignments.
  \end{minipage}

\end{center}
\vspace{1em}


\section{Introduction}

Genome rearrangement plays a fundamental role in biological processes
including cancer \cite{greenman12estimation}, gene
regulation, and development \cite{nowacki11rna} and 
a better understanding of genome rearrangement is ex\-pect\-ed to lend
insight into these biological processes.
The primary evidence for genome rearrangement in modern genomes comes from identifying rearrangement breakpoints in alignments of two or more genomes.
Genome alignments and the rearrangement breakpoints they encode provide a basis for
reconstructing genome rearrangement histories.
The accurate reconstruction of rearrangement history depends intimately on
whether genome alignments can accurately identify breakpoints.

When genomes have undergone gene loss in addition to rearrangement some breakpoints may only be detectable by comparison of three or more genomes. This class of breakpoints has received limited attention to date.
We refer to such breakpoints as ``hidden'' breakpoints, since usage of these rearrangement breakpoints in an organism's evolutionary history is not apparent from pairwise comparison of available genomes.
Hidden breakpoints occur in regions of the genome conserved in some, but not all of the genomes under comparison.
A simplest possible example of hidden breakpoints involves the three genomes $A$, $B$, $C$, with blocks 1, 2, 3 (see Fig.\,\ref{fig:3way}).

By extending a recent solution to the
breakpoint median problem \cite{tannier09multichromosomal}, we demonstrate that
it is possible to calculate a hidden breakpoint count in
three genomes when there is gain and loss. 
Our method is founded on the premise that a parsimonious reconstruction of the median genome would contain any sequence content present in at least two of the three genomes, while sequence content unique to a single genome would not be in the median.
By comparing pairwise distances to the median distance, our method reveals some of the hidden breakpoints.
Having implemented the method to calculate hidden breakpoint counts, we then conduct a simulation study to measure the abundance of hidden
breakpoints under a range of different genome evolution parameters.
We apply current multiple genome alignment systems to
calculate alignments of the simulated sequences.
Finally, we count hidden breakpoints in the calculated
alignments and compare with the true number of hidden breakpoints in
the simulated sequences. 
We show that all tested genome alignment algorithms introduce error in hidden breakpoint estimation and characterize their error rates under a range of evolutionary parameters in simulation.


\section{Pairwise and Hidden Breakpoints\label{sec:breakpoints}}

Genome alignments consist of sequence regions that are aligned in two or more genomes. 
Each aligned sequence region, often called a synteny block or locally collinear block, is internally free from any rearrangement but may contain gaps.
Here we simply call these regions \emph{blocks}.
We assign an integer-valued identifier to each block of an alignment, using the sign of the integer to represent whether the block occurs in a genome in the forward or in reverse orientation.
In this way, the arrangement of a single genome $g$ in a genome alignment can be represented as a sequence of signed integers: $g = b_1\ b_2\ \dots\ b_n$, $b_i \in \mathbb{Z}$.
Note that this sequence of integers does not necessarily represent all parts of the genome, i.\,e. there may be unaligned, unique segments in-between blocks.
Genomes themselves may consist of one or multiple chromosomes with either linear or circular topology. 
When a genome contains multiple chromosomes, it cannot be represented as a single sequence of integers, but rather as a set of sequences: $g = \{b_1 \dots b_i, b_{i+1} \dots b_{i+j}, \dots\}$.
Following previous literature we define linear chromosomes to be bounded by zero length \emph{telomeres}, acknowledging the discrepancy between this definition and its use in the biological literature to describe a string of nucleotides of length $> 0$ near the ends of a linear chromosome.
We use $\bullet$ to denote telomeres in a linear chromosome: $\bullet\ b_1\ \dots\ b_n\ \bullet$.

\begin{figure}
\centerline{\includegraphics[scale=0.67]{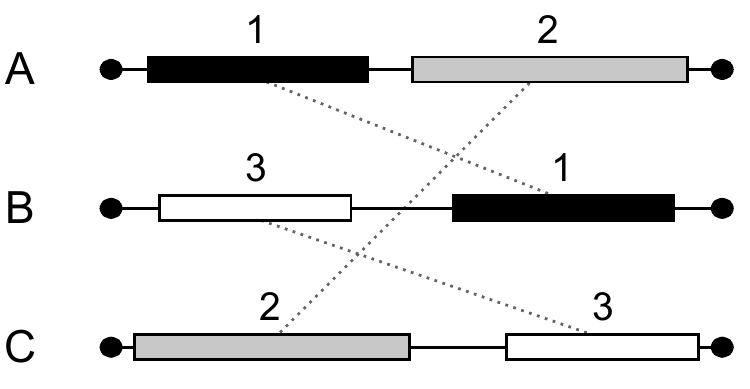}\hfill
            \includegraphics[scale=0.67]{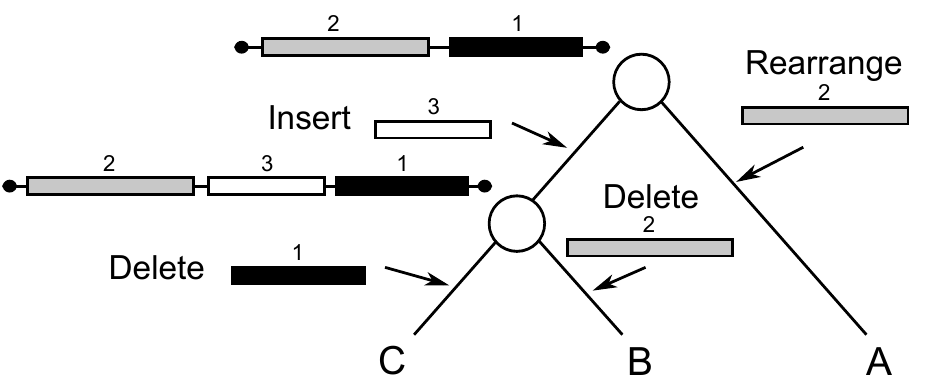}}
\caption{Multiple alignment (left) and possible evolutionary scenario (right) of genomes A, B, and C containing a rearrangement not detectable in pairwise comparison. In all three pairwise projections two of the blocks become unique leading to a breakpoint distance of 0.}
\label{fig:3way}
\end{figure}
In the present work we investigate rearrangement breakpoints in groups of two or three genomes, although an alignment might contain many more genomes $g_1, \dots, g_m$.
This requires us to define a \emph{projection} of an alignment to those blocks present in the current set of genomes.
For three genomes $g_x$, $g_y$, and $g_z$ with $x,y,z \in \{1, \dots, m\}$ we denote the projection of the alignment as $\pi_{xyz}$, and a genome in the projected alignment by $\pi_{xyz}(g)$.
We define the projected genome $\pi_{xyz}(g_x)$ as the subsequence of $g_x$ containing those blocks present in at least two of the genomes used for projection.
We thus remove blocks that are unique to $g_x$, $g_y$, or $g_z$ from consideration in the three-way comparison since they can not reveal direct evidence of rearrangement breakpoints.

Two blocks $b_i, b_j$ that are consecutive within a genome (possibly with unique segments in-between) define an \emph{adjacency}.
In a pairwise projection of a genome alignment, a \emph{pairwise breakpoint} is an adjacency from one genome that is missing in the other.
By counting breakpoints in one genome $g_x$ relative to another $g_y$ we can calculate the \emph{breakpoint distance} $d(g_x,g_y)$.
Breakpoints at telomeres count only $\frac{1}{2}$ in $d(g_x,g_y)$, as only breakage and no fusion to another loose end has happened.
When $g_x$ and $g_y$ have equal content $d(g_x,g_y)$ is equivalent to the classic breakpoint distance.
If different sets of blocks are present in $g_x$ and $g_y$, the breakpoint distance may be asymmetric, i.\,e. $d(g_x,g_y) \neq d(g_y,g_x)$.
For brevity we define $d_{xyz}(g_x, g_y) := d(\pi_{xyz}(g_x), \pi_{xyz}(g_y))$.

Previous work has applied pairwise breakpoint distance in a sum-of-pairs score for multiple genome alignment \cite{darling10progressiveMauve}.
However, multiple alignments may reveal rearrangements that are not visible in pairwise projections (see example in Fig.\,\ref{fig:3way}), information that sum-of-pairs scores do not capture.
The breakage events of these rearrangements are hidden by gain or loss of rearranged blocks or by breakpoint re-use.
We thus use the term \emph{hidden breakpoint} to refer to this class of breakpoints.


\section{The Median Approach for Counting Breakpoints\label{sec:counting}}

We now describe a method to compute a hidden breakpoint count $\mathcal{H}$ for three genomes.
For alignments with $m > 3$ genomes, $\mathcal{H}$ can be calculated for all projections to three genomes.

\begin{figure}
\centerline{\includegraphics[scale=0.6]{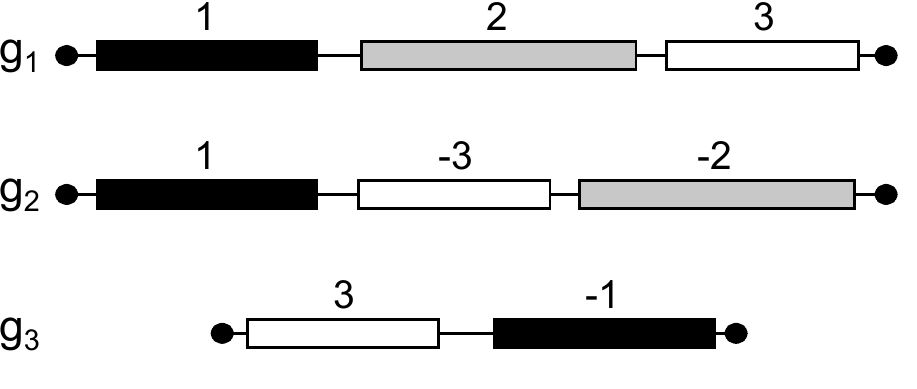}\hfill
            \includegraphics[scale=0.6]{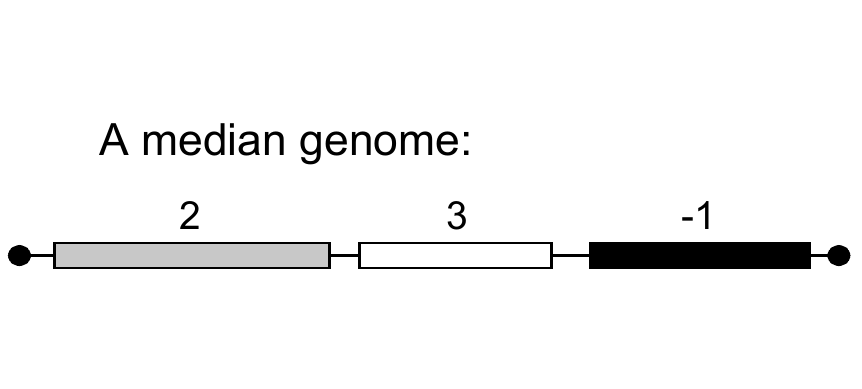}}
\caption{Alignment of three genomes and a corresponding median genome. The pairwise distances to the median $M$ are $d(g_1,M)=1.5$, $d(g_2,M)=0$, and $d(g_3,M)=0.5$.} 
\label{fig:median}
\end{figure}

A median $M$ for three genomes $g_x$, $g_y$, $g_z$ with $x,y,z\in\{1, \dots, m\}$ is a genome formed on all blocks in $\pi_{xyz}$.
The blocks in $M$ are arranged such that the sum 
\[
d_M := \sum_{k \in \{x,y,z\}} d_{xyz}(g_k, M)
\]
is minimized (see Fig.\,\ref{fig:median} for an example).
This represents a generalization of previous definitions of a median genome, e.\,g.~\cite{tannier09multichromosomal}, to include all blocks in at least two out of three genomes $g_x, g_y, g_z$. Previous breakpoint median methods required blocks to be conserved among \emph{all} genomes.
As we describe in Section~\ref{sec:counts}, removing the portion of $d_M$ attributed to pairwise breakpoints yields a hidden breakpoint count.
We note that our method only requires computation of $d_M$ and does not depend on accurate reconstruction of the actual median genome, for which many possibilities with minimal $d_M$ may exist.
We also note that duplications are forbidden; each block may occur at most once per genome.

We compute the median distance $d_M$ by applying a recent polynomial time solution~\cite{tannier09multichromosomal} to our generalization of the median definition.
Following the proof of Theorem~1 in~\cite{tannier09multichromosomal}, we construct a graph from the genomes (see Sect.\,\ref{sec:graph} for details), and compute a maximum weight perfect matching which corresponds to a median genome~\cite{tannier09multichromosomal}.
Because counting hidden breakpoints requires only the median distance $d_M$ and not the actual median, we are able to improve the time complexity of the solution from \cite{tannier09multichromosomal} by reducing the number of graph edges (see Sect.\,\ref{sec:edges}) input to the perfect matching.
Note that this reduction eliminates some valid medians, but the median distance $d_M$ is unchanged.

\subsection{Graph Construction.\label{sec:graph}}

\begin{figure}
\centerline{\includegraphics[width=\linewidth]{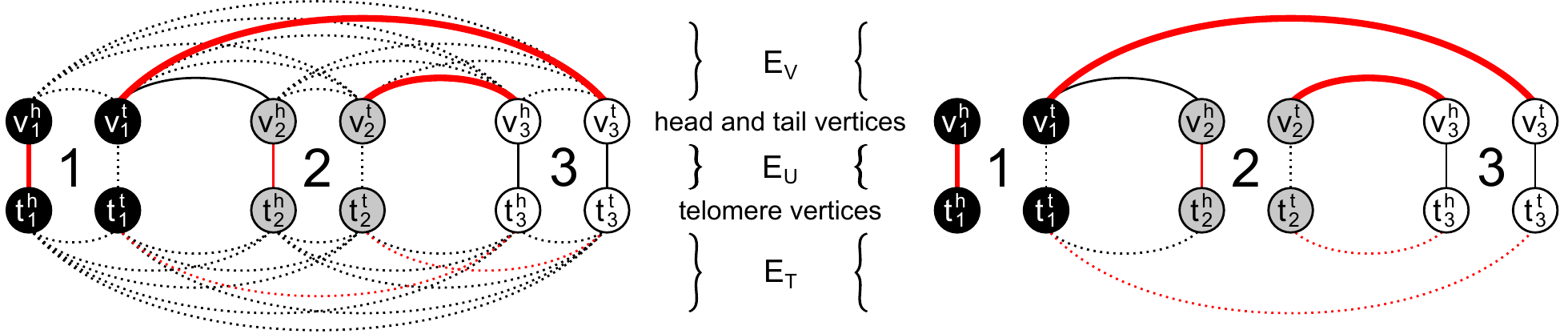}}
\caption{Full graph (left) and a simplified graph (right) for the example alignment from Fig.\,\ref{fig:median}.
The simplified graph is sufficient to calculate the median distance $d_M$.
Line width indicates edge weights.
Dotted edges have a weight of 0.
Red edges form a maximum weight perfect matching that corresponds to the median genome shown in Fig.\,\ref{fig:median}.}
\label{fig:matchingGraph}
\end{figure}

We first describe the graph as used in \cite{tannier09multichromosomal} to compute a median genome.
Figure\,\ref{fig:matchingGraph} (left) shows an example graph for the alignment from Fig.\,\ref{fig:median}.
Given a genome alignment, the graph $G=(V\cup T,E_V\cup E_T\cup E_U)$ contains four vertices from two different vertex sets V and T for each block $b_i$ of the alignment:
A \emph{head} vertex $v_i^h \in V$ representing the start of the block, a \emph{tail} vertex $v_i^t \in V$ representing the end of the block, a \emph{head telomere} vertex $t_i^h \in T$, and a \emph{tail telomere} vertex $t_i^t \in T$.
The telomere vertices $t_i^h$ and $t_i^t$ represent the possibilities that chromosomes start or end with block $b_i$.
It is necessary to have separate telomere vertices for each block to be able to find all (possibly multichromosomal) medians by the perfect matching computation.
All vertices from $V$ are connected among each other by edges $e_V \in E_V$, and all vertices from $T$ are connected among each other by edges $e_T \in E_T$, i.\,e. the two subgraphs are complete.
In addition, each head vertex $v_i^h$ is connected to its corresponding head telomere vertex $t_i^h$, and each tail vertex $v_i^t$ to its corresponding tail telomere vertex $t_i^t$ by edges $e_U \in E_U$.
Initially, all edges have a weight of 0.
We increase edge weights according to adjacencies in the genomes.
For telomere adjacencies, we increase weights of edges in $E_U$ by $\frac{1}{2}$ following \cite{tannier09multichromosomal}.
For all other adjacencies we increase weights of edges from $E_V$ by 1.
Depending on the orientation (sign) of the adjacent blocks, we increase the weight of the edge between a tail and head vertex, tail and tail, head and tail, or head and head vertex by 1.
All edges in $E_T$ keep a weight of 0.

\subsection{Reducing the Number of Edges.\label{sec:edges}}

Having constructed the graph, we compute a maximum weight perfect matching.
The running time of the most efficient maximum weight perfect matching algorithms (e.\,g.~\cite{kolmogorov09blossom}) depends on the number of edges in the graph, which in our case is quadratic in the number of blocks.
In the following we show how it is possible to reduce the number of edges by a linear factor.

A perfect matching is a subset of the graph's edges $E_M \subseteq E_V \cup E_T \cup E_U$ such that each vertex of the graph is incident to exactly one edge in the matching.
A maximum weight perfect matching is the perfect matching that maximizes the sum of weights of all edges $e \in E_M$.
Figure\,\ref{fig:matchingGraph} (left) demonstrates that many of the edges in a graph constructed as described in \ref{sec:graph} have a weight of 0, and hence, will not contribute to the weight of a maximum weight perfect matching.
Still, some of those edges are necessary to allow a perfect matching of maximum weight, but not all of them.
In the following paragraph we explain why edges with a weight of 0 in $E_V$ between vertices $v_i^p$ and $v_j^q$, with $p,q \in \{h,t\}$ and with $b_i,b_j$ being blocks, can be removed from the graph as well as corresponding edges in $E_T$ between vertices $t_i^p$ and $t_j^q$ without affecting the weight of the resulting maximum weight perfect matching.

A graph that was constructed as described above can have multiple maximum weight perfect matchings.
Given any of these matchings, we can replace the set of edges $E_M \cap E_T$ by those edges between vertices $t_i^p, t_j^q \in T$, for which the matching contains an edge between $v_i^p, v_j^q \in V$.
This replacement does not violate the matching property and does not affect the weight of the matching since all edges in $E_T$ have a weight of 0.
The matching may then contain some pairs of 0-weight edges from $E_V$ and $E_T$ connecting vertices $v_i^p, v_j^q \in V$ and connecting vertices $t_i^p, t_j^q \in T$. We can replace such pairs of edges by edges from $E_U$ between $v_i^p$ and $t_i^p$ and between $v_j^q$ and $t_j^q$, again without violating the matching property and affecting the weight of the matching.
After this replacement none of the 0-weight edges from $E_V$ and the corresponding edges from $E_T$ are part of the matching.
Therefore, a maximum weight perfect matching exists that does not use these pairs of edges, and we can remove them from the initial graph without affecting the weight of the resulting maximum weight perfect matching.

\subsection{Hidden Breakpoint Counts.\label{sec:counts}}

A maximum weight perfect matching on the above described graph assigns each head or tail vertex to its telomere vertex or to another head or tail vertex by edges.
These edges correspond to the adjacencies of a median genome.
The edges that are not part of the matching but have a weight greater than 0, correspond to breakpoints among the median and one or more other genomes.
Thus, by calculating the weight difference of the original graph and the matching, we obtain the total number of breakpoints $d_M$ between the three original genomes $g_x$, $g_y$, and $g_z$ and the median genome $M$.
However, $d_M$ overestimates the actual number of breakpoints among the three genomes.
The median genome may contain blocks that are not present in one of the genomes, which disrupt adjacencies of blocks that occur in the same order without rearrangement.
These disrupted adjacencies are neither pairwise nor hidden rearrangement breakpoints.
Therefore, we calculate a separate hidden breakpoint count
\[
\mathcal{H} = d_M - f \enspace .
\]
The second term $f$ removes the fraction of $d_M$ which is due to breakpoints observable in pairs of genomes including those disrupted adjacencies caused by unequal block content. $f$ is defined as:
\[
f = \frac{1}{2} \sum_{a<b\in\{x,y,z\}} d_{xyz}(g_a,M_{ab}) + d_{xyz}(g_b,M_{ab}) \enspace .
\]
For the computation of $f$ we use a median $M_{xy}$ of two genomes $g_x$, $g_y$, which is formed on all blocks of the triplet projection $\pi_{xyz}$.
We can compute $M_{xy}$ using the same approach as for a median of three genomes, i.\,e. construction of the graph and computation of a maximum weight perfect matching.
The multiplication by $\frac{1}{2}$ is required because the distance from each genome to a median is counted exactly twice (to different medians).

In the results section we discuss hidden breakpoints in alignments containing arbitrarily many genomes. 
In alignments with more than three genomes, we calculate $\mathcal{H}$ for all projections to three genomes and report the sum over all projections.
Likewise, for pairwise breakpoints we report the sum over all breakpoints in pairwise projections.

\section{Simulating Evolution\label{sec:simulation}}

We simulated genome evolution to generate sets of evolved sequences and genome alignments relating them using the previously described method sgEvolver~\cite{darling04mauve}.
Briefly, sgEvolver applies the standard Markov process model of evolution (spe\-cif\-i\-cal\-ly a marked Poisson process), with mutation events including substitutions, insertions and deletions of three different size distributions, and inversions. 
In our simulation each of the mutation types has the following properties:
\begin{description}
\item{\bf Nucleotide substitutions} follow an HKY process with a tran\-si\-tion/trans\-ver\-sion ratio of 4 and background nucleotide frequencies $\mbox{A}=0.265$, $\mbox{C}=0.235$, $\mbox{T}=0.265$, $\mbox{G}=0.235$, as implemented in the program seq-gen~\cite{rambaut97seqgen}.
Gamma-distributed heterogeneity in substitution rates across sites was simulated with the shape parameter $k = 1$.
\item{\bf Indels} comprise the first class of insertions and deletions. Indel lengths follow a Poisson distribution with $\lambda = 3$.
\item{\bf Small gain/loss} events comprise the second insertion/deletion class and have lengths geometrically distributed with $p = 200$.
\item{\bf Large gain/loss} events comprise the final size class of insertions and deletions and have lengths uniformly distributed between 10000 and 50000.
\item{\bf Inversions} follow a geometric length distribution with $p = 50000$.
\end{description}
These events were simulated on a phylogeny containing nine extant taxa.
All insertion/deletion events are simulated to have uniformly random distributed positions in the genome.
The rate of each event type is specified by a relative rate parameter; the expected event count per site on a particular branch of a phylogeny is the product of branch length and relative rate.
The ancestral genome in our simulations is 1\,Mbp in size, and since insertions and deletions occur with equal probability, the expected genome size remains constant throughout the simulation process (though the variance grows).

The approach used by sgEvolver to simulate insertions can also introduce duplications even though they are not directly modeled.
In order to simulate insertions, sgEvolver maintains a finite pool of ``donor'' genomic material.
When an insertion occurs, the sequence to insert is sampled uniformly at random from the donor material pool.
This approach is intended to model the total pool of gene content available in a population, sometimes called a pan-genome~\cite{medini05pangenome}.
When the simulation includes a large enough number of insertion events, especially ``large gain/loss'', it becomes likely that the same material from the donor pool will be inserted more than once, effectively creating a duplication in the simulated genome.
sgEvolver does not keep track of the donor pool positions for insertion, and therefore the simulated genome alignments ignore possible duplications.

When simulating 1\,Mbp genomes, the simulation and subsequent genome alignment process is fast enough that we can analyze a large range of parameters in parallel on a compute cluster.
We report results on various parameter ranges below.


\section{Hidden Breakpoints in Simulated Data\label{sec:results}}

In this section we examine the effect of inversion, gain/loss, and nucleotide substitution rate on pairwise and hidden breakpoints.
We expect hidden breakpoints to be driven by inversion and gain/loss, while nucleotide substitution may affect breakpoint counts in calculated alignments by increasing the difficulty of accurate alignment.

We describe two simulated data sets below, \emph{InvNt} and \emph{InvGL}, consisting of 400 and 200 genome alignments respectively.
On each alignment of the datasets we computed the true number of pairwise breakpoints and hidden triplet breakpoints (see Sect.\,\ref{sec:resultsTrue}).
Then in Sect.\,\ref{sec:resultsAligners}, we present calculated alignments of the simulated sequences done with three different genome alignment systems.
We discuss how calculated alignments generally overestimate breakpoint counts and relate this error to evolutionary rates and other measures of alignment accuracy.

\subsection{True Breakpoint Counts in Simulated Genome Alignments.\label{sec:resultsTrue}}

\begin{figure}
\centerline{\includegraphics[width=0.45\linewidth]{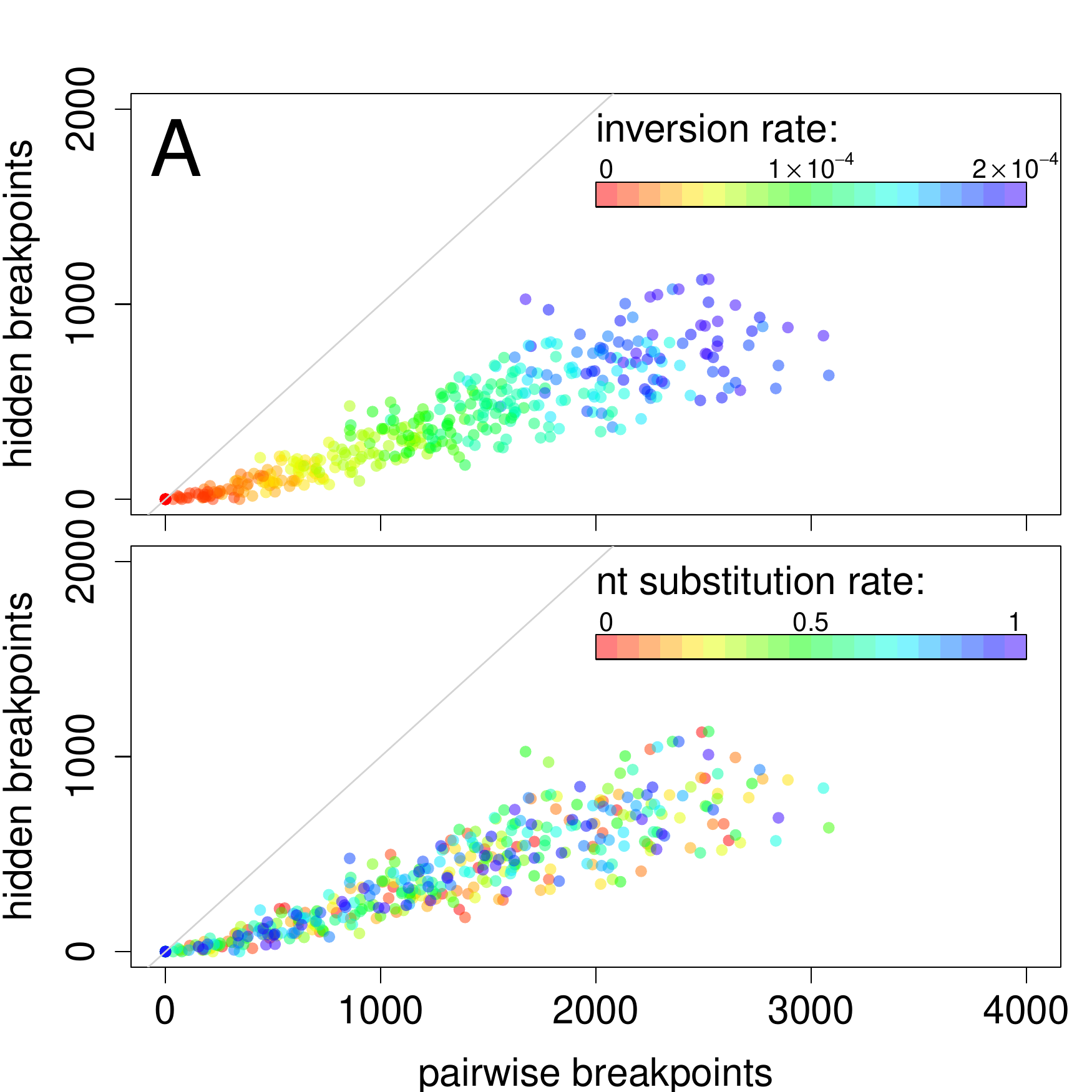}\hfill
            \includegraphics[width=0.45\linewidth]{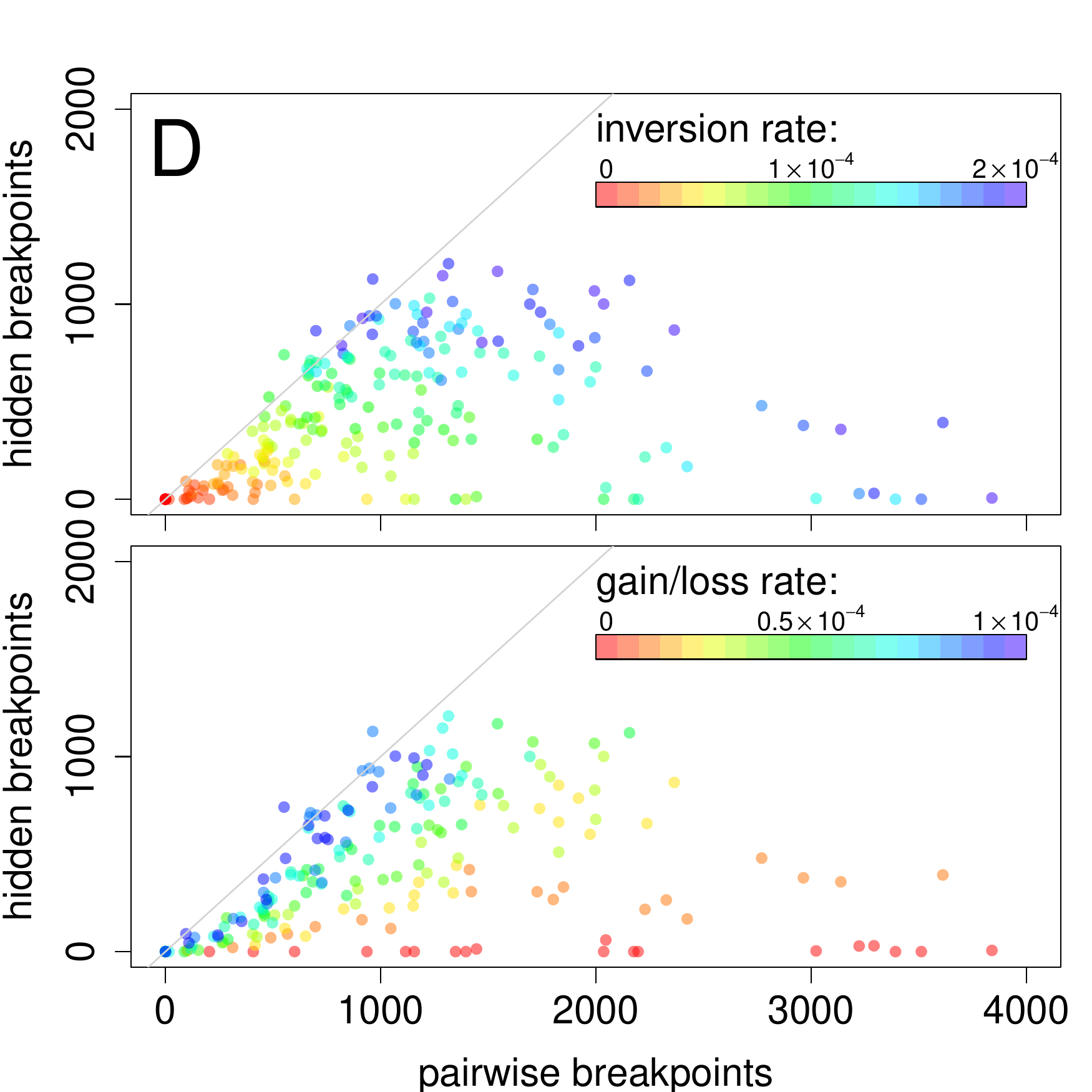}}
\centerline{\includegraphics[width=0.45\linewidth]{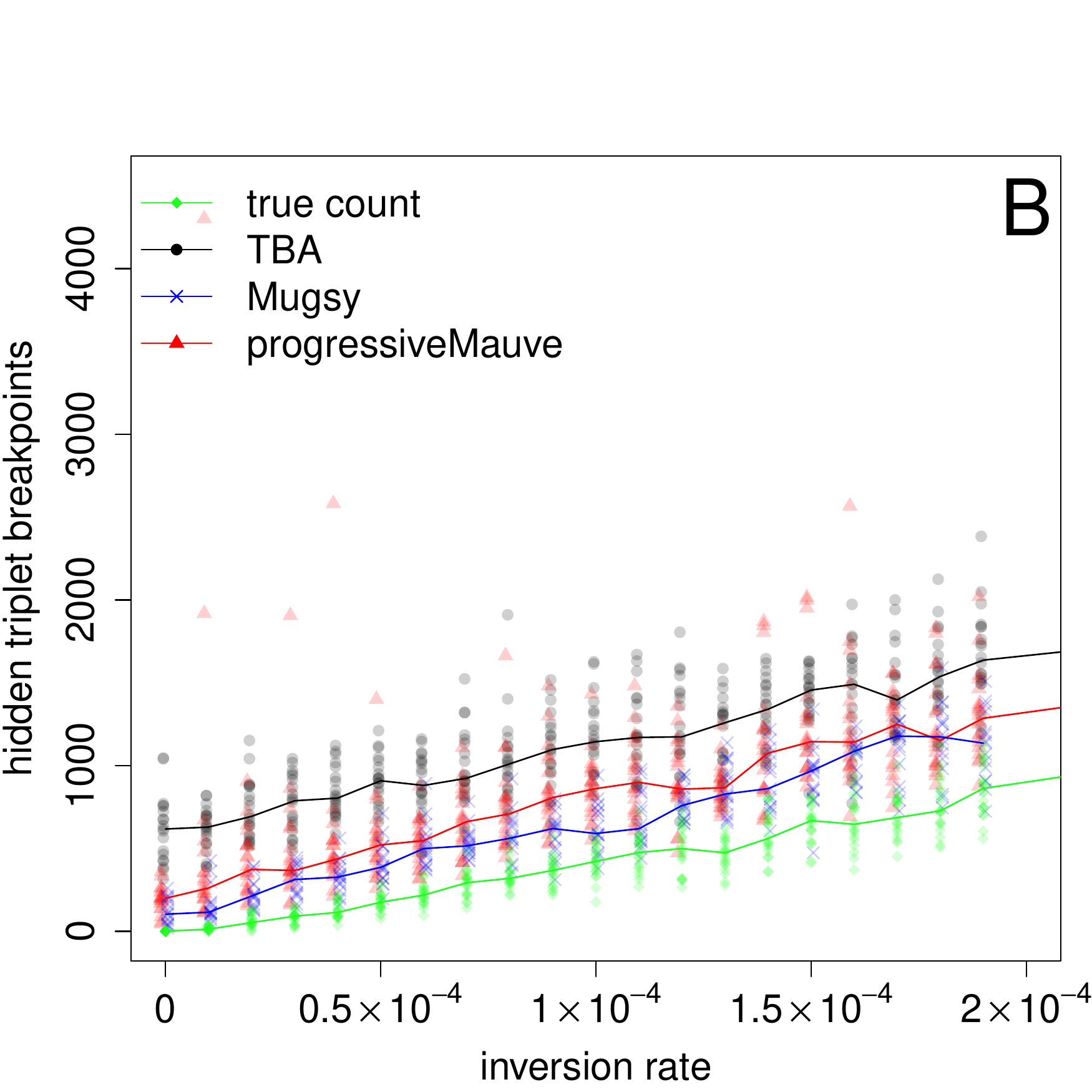}\hfill
            \includegraphics[width=0.45\linewidth]{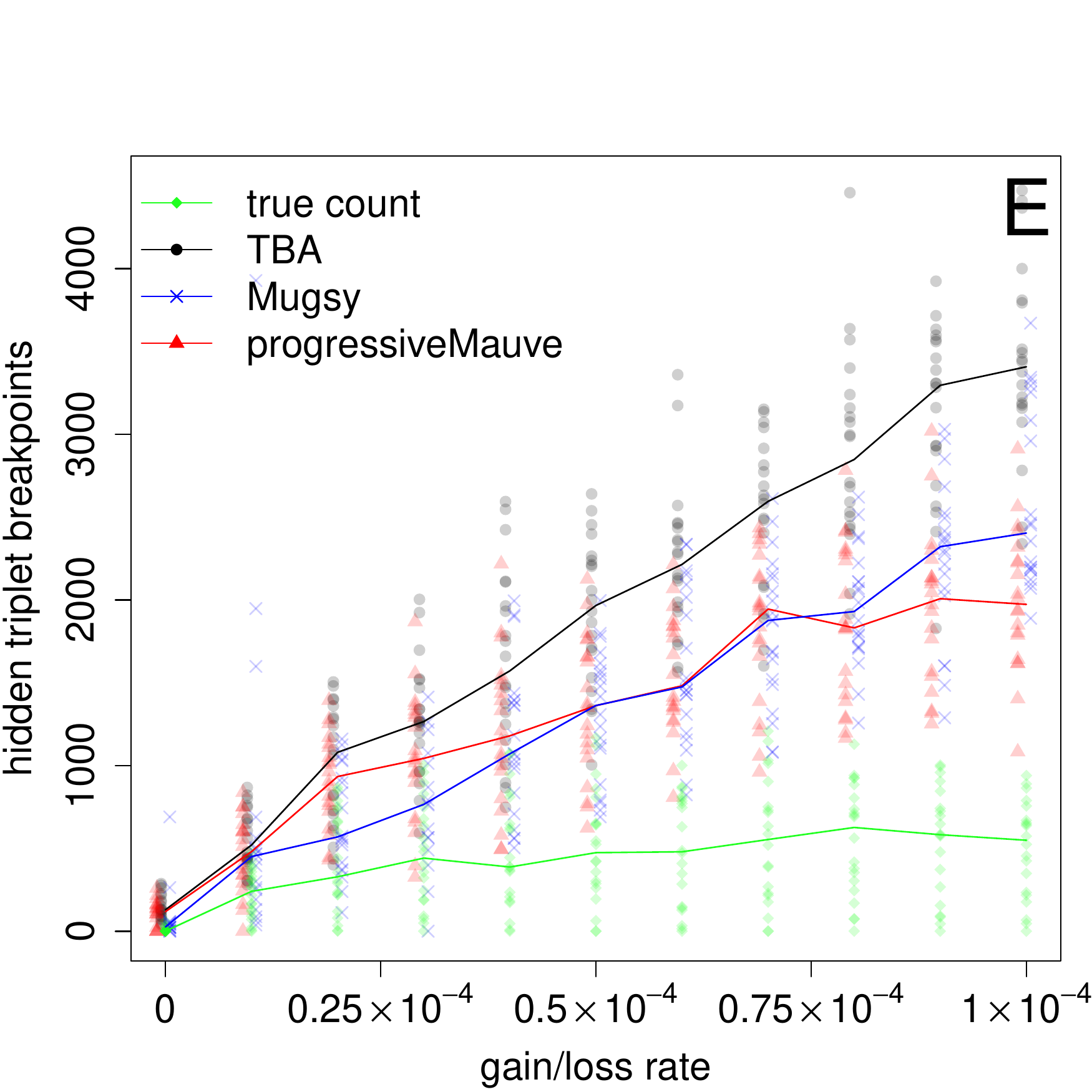}}
\centerline{\includegraphics[width=0.45\linewidth]{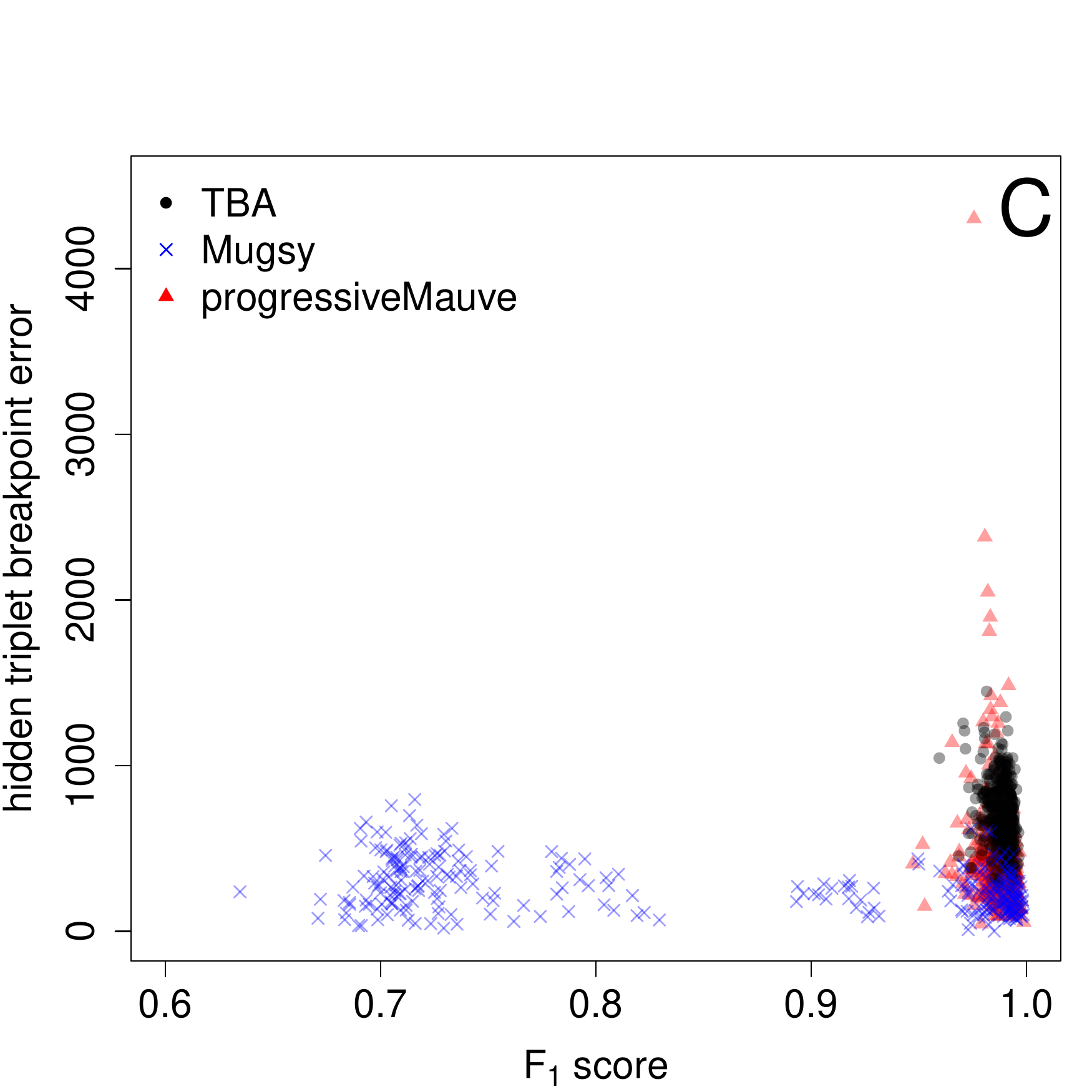}\hfill
            \includegraphics[width=0.45\linewidth]{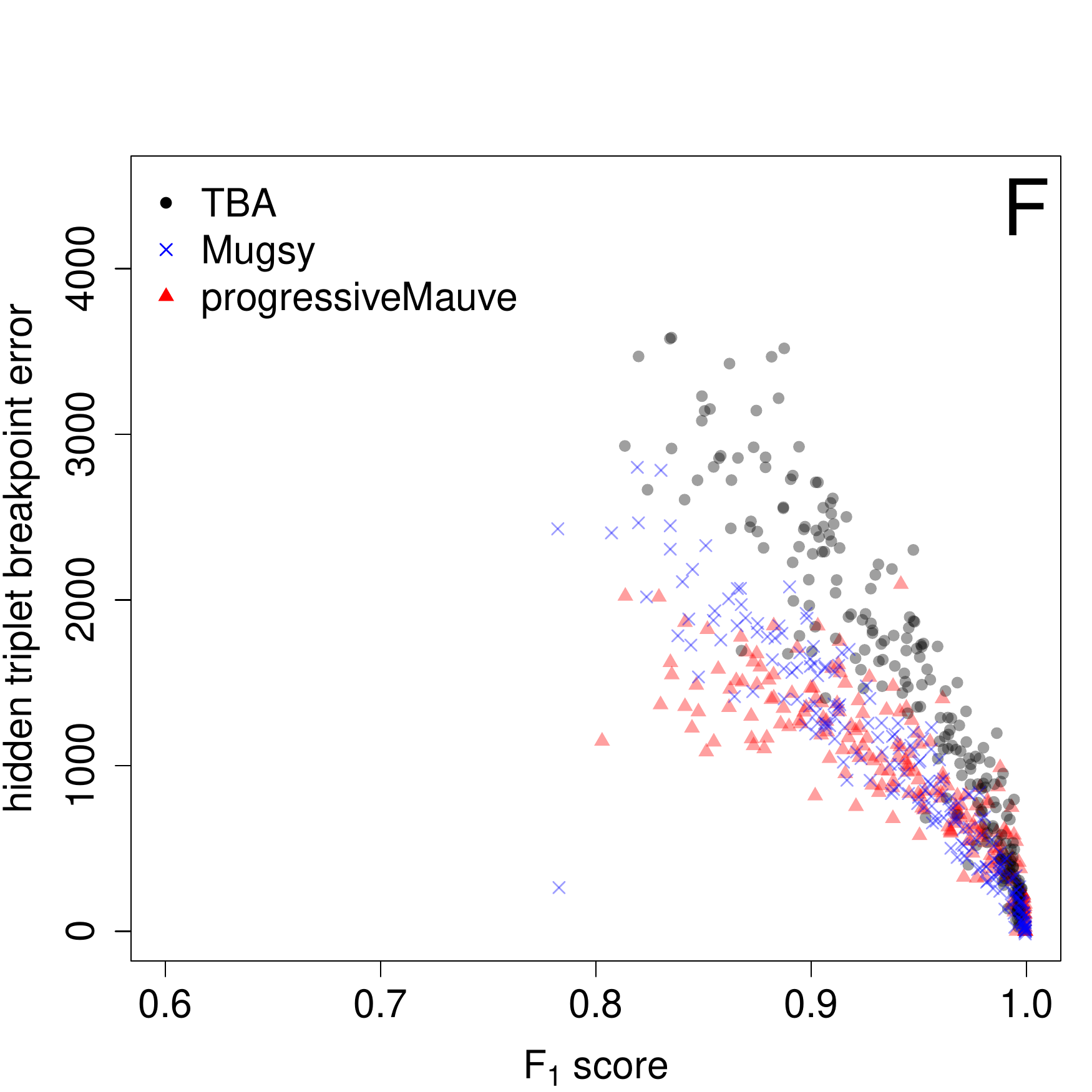}}
\caption{Hidden triplet breakpoint counts in simulated alignments with different inversion and nucleotide substitution rates (\textit{InvNt}, A--C), and 200 simulated alignments with different inversion and large gain/loss rates (\textit{InvGL}, D--F).}
\label{fig:plots}
\end{figure}

The \emph{InvNt} data set enables examination of the influence of the inversion and nucleotide substitution rate on the breakpoint counts.
We expect nucleotide substitutions to have no effect on breakpoint counts in simulated alignments, although as we discuss later in calculated alignments a high substitution rate affects estimated breakpoint counts by making alignment difficult.
We simulated 400 alignments with inversion rates between 0 and $2\times 10^{-4}$ in steps of $1\times 10^{-5}$ and nucleotide substitution rates between 0 and 1 in steps of 0.05.
All three insertion/deletion rates were fixed: indels at $1 \times 10^{-3}$, small gain/loss at $5 \times 10^{-4}$, and large gain/loss at $2 \times 10^{-5}$.
Previous work has demonstrated that alignment algorithms can reconstruct highly accurate genome alignments at these relatively low insertion/deletion rates~\cite{darling10progressiveMauve}. 
In the alignments we observe 0 to 3100 pairwise breakpoints and 0 to 1150 hidden triplet breakpoints (see Fig.\,\ref{fig:plots}A).
Despite these large ranges, the ratio between pairwise and hidden breakpoints stays constant.
The coloring in Fig.\,\ref{fig:plots}A (top) shows that with growing inversion rates, we obtain more breakpoints, both pairwise and hidden triplet breakpoints.
As one would expect, the nucleotide substitution rate has no direct effect on the number of breakpoints (Fig.\,\ref{fig:plots}A, bottom).

In the \emph{InvGL} data set we examine breakpoint counts for different large gain/loss and different inversion rates.
\emph{InvGL} consists of 200 alignments with large gain/loss rates between 0 and $1 \times 10^{-4}$ step size $1\times 10^{-5}$ and again inversion rates between 0 and $2\times 10^{-4}$ in steps of $1\times 10^{-5}$.
The other two insertion/deletion rates were fixed at the same values as above, and the nucleotide substitution rate at 0.01.
As with \emph{InvNt}, we observe more pairwise and hidden triplet breakpoints for growing inversion rates (Fig.\,\ref{fig:plots}D, top), but in contrast to \emph{InvNt}, the ratio between pairwise and hidden breakpoint counts varies.
The coloring in Fig.\,\ref{fig:plots}D (bottom) demonstrates that this variation is due to different gain/loss rates:
With growing gain/loss rate, we observe more hidden triplet breakpoints whereas the number of pairwise breakpoints goes down slightly.
This result is consistent with our definition of hidden breakpoints.
The slight decrease in pairwise breakpoints at higher gain/loss rates is due to removal of breakpoints by gene loss.
The hidden breakpoint count is also affected by gene loss in the \textit{InvGL} simulation, reaching a maximum at a gain/loss rate of about 0.8.
%

\subsection{Tested Alignment Programs and Evaluation Metrics.\label{sec:aligners}}

Having established the true counts of hidden triplet breakpoints under a range of evolutionary scenarios, we continue by evaluating whether some current algorithms can infer genome alignments with the correct number of hidden breakpoints. 
The algorithm to calculate hidden triplet breakpoints requires positional homology alignments, the method cannot be applied to genome alignments containing paralog alignments.
For this reason we selected and applied three current programs that generate positional homology genome alignments: TBA, progressiveMauve, and Mugsy.

The \textbf{Threaded Blockset Aligner (TBA)}~\cite{blanchette04aligning} constructs multiple genome alignments by a process of merging and filtering pairwise alignments generated by BLASTZ~\cite{schwartz03human}. 
TBA requires specification of a phylogenetic guide tree, and for this we gave it the same topology used for simulation. 
We used TBA version 2009-Jan-21, the most recent version publicly available at the time of this work.
The \textbf{progressiveMauve} genome alignment algorithm~\cite{darling10progressiveMauve} begins by identifying approximate multi-matches among input genomes, then progressively grouping these into locally collinear blocks and aligning nucleotides within these blocks. 
For grouping matches progressiveMauve applies a sum-of-pairs breakpoint scoring scheme, which penalizes pairwise breakpoints.
In the present work we used progressiveMauve version 2011-02-02 with default options.
\textbf{Mugsy} is a newer aligner~\cite{angiuoli11mugsy} that also uses three steps as progressiveMauve, but differs in the details of each step.
For example, it uses a min-cut max-flow algorithm to partition the multi-matches into blocks under a synteny score.
We used Mugsy v1r2.2 with default options.

For each alignment from the three aligners we computed the sum of hidden triplet breakpoint counts $\mathcal{H}$.
To assess the error that aligners introduce to alignments in terms of breakpoints, we calculated the difference between the number of breakpoints in calculated alignments and the number of breakpoints in the true alignments.
We also calculated a score that measures nucleotide alignment accuracy independently of genome arrangement:
We compared each pair of aligned nucleotides in the calculated alignments with the true alignment to obtain precision and recall of nucleotide positional homology prediction, and combined the two values to an F$_1$ score.

\subsection{Breakpoint Counts in Calculated Genome Alignments.\label{sec:resultsAligners}}

We computed alignments with TBA, progressiveMauve, and Mugsy on all 600 sets of nine evolved genomes in the two data sets \emph{InvNt} and \emph{InvGL}.
Figures~\ref{fig:plots}B and \ref{fig:plots}C show hidden triplet breakpoint counts of the alignments from \emph{InvNt}, and Fig.\,\ref{fig:plots}E and \ref{fig:plots}F for \emph{InvGL}.
In Fig.\,\ref{fig:plots}B and \ref{fig:plots}E, we display the median of the hidden triplet breakpoint counts per program for alignments simulated with equal inversion (\ref{fig:plots}B) or gain/loss rate (\ref{fig:plots}E) as lines.

In alignments calculated on \emph{InvNt} we found that all three programs overestimate the hidden breakpoint counts by a constant value independent of the inversion rate, suggesting a relationship with the constant gain/loss rate in this dataset.
The two aligners that use an algorithm designed to minimize pairwise breakpoints introduce fewer additional hidden triplet breakpoints than TBA.
The F$_1$ score of most alignments is very good ($>0.95$) except for some Mugsy alignments.
We recognize a trend that alignments with F$_1$ scores closer to 1 have lower hidden triplet breakpoint errors.
An exception are Mugsy alignments on sequences simulated with nucleotide substitution rates greater than 0.4.
These Mugsy alignments have a very low recall suggesting the simulated genomes were too divergent to be aligned by Mugsy with its default settings.

Calculated alignments of \emph{InvGL} data show that the amount of gene gain/loss has a strong effect on the hidden triplet breakpoint error.
The error is small at low rates of gain/loss, but grows quickly with the gain/loss rate.
In Figure\,\ref{fig:plots}F we see that accurate alignments as measured by F$_1$ have lower hidden triplet breakpoint error.
We find that the F$_1$ score drops with increasing gain/loss rates (data not shown), suggesting that genomes with high rates of simulated gain/loss may be generally difficult to align.


\section{Conclusions}

We have introduced the concept of hidden breakpoints in genome alignments and presented a method to calculate hidden triplet breakpoint counts.
We have examined hidden breakpoint counts in simulated alignments evolved over a wide range of evolutionary parameters,
and in alignments calculated by three different genome alignment algorithms on the same simulated datasets.
We find that all tested genome aligners introduce a high degree of error 
in hidden breakpoint counts compared to the true count, and that this error grows with
evolutionary distance in the simulation. 
Our results suggest that hidden breakpoint error may be pervasive in
genome alignments.
The error may in turn lead to erroneous inference of ancestral genomes and rearrangement history.
Therefore, studies of the relationship between genome
rearrangement history and biological processes
could be improved by considering this error during the computation of genome alignments.

Overall, the error in triplet breakpoint counts is of the same order of magnitude as the pairwise breakpoint error.
Methods such as the sum-of-pairs breakpoint score in progressiveMauve reduce the pairwise breakpoint error, but may not effectively reduce hidden breakpoint error.
The max-flow min-cut algorithm employed by Mugsy appears to yield the most precise block and breakpoint estimates, but only when substitution rates are low enough.
It is possible that the use of a more sensitive matching method such as promer~\cite{delcher02fast} or another approximate matching approach would enable Mugsy to produce high accuracy results also at higher levels of sequence divergence. 
Future alignment systems would benefit from approaches that either explicitly or implicitly consider hidden breakpoints.

\paragraph{Limitations.}
The presented concept of hidden breakpoints has some limitations.
First, the pairwise breakpoint distance is a lower bound for the number of breakage events that happened during evolution because several rearrangement events may use the same breakpoint.
The hidden breakpoint count described here improves the estimate of breakage events, but it remains a lower bound to the actual number of breakage events that happened during evolution.
A further limitation is that the current approach does not directly give rise to a method for phylogenetic inference on genome arrangement.
Others have demonstrated practical algorithms for phylogenetic inference on genome arrangement using DCJ medians with gene gain and loss~\cite{zhang10phylogenetic}. 
Although our median algorithm could be used in such a context, there is no constraint on the number or topology of chromosomes in median genomes and this might yield inferred ancestral genomes that are extremely biologically unlikely.
Another limitation is the perfect-matching based median approaches cannot handle duplicated blocks in the alignments.
Finally, we have only derived a formula to calculate hidden breakpoints among three genomes but some hidden breakpoints may be identifiable only among 4 or more genomes.
We leave identification of hidden breakpoints in 4 or more genomes, and analysis of genomes with duplications as future work.


\subsubsection*{Acknowledgements.}
This work was enabled by funding from the DFG through the Dahlem Research School.
We have implemented the calculation of breakpoint counts using the SeqAn C++ library (http://www.seqan.de), and the maximum weight perfect matching implementation from the Lemon graph library (http://lemon.cs.elte.hu).


\small
\bibliography{hidden_breakpoints}

\end{document}